\documentclass{PoS}

\def\mathswitchr#1{\relax\ifmmode{\mathrm{#1}}\else$\mathrm{#1}$\fi}
\newcommand{\PH}{\mathswitchr H}
\newcommand{\Pp}{\mathswitchr p}
\newcommand{\Pb}{\mathswitchr b}
\newcommand{\Pt}{\mathswitchr t}
\newcommand{\Pu}{\mathswitchr u}
\newcommand{\Pd}{\mathswitchr d}
\newcommand{\Pg}{\mathswitchr g}
\newcommand{\PW}{\mathswitchr W}
\newcommand{\Pw}{\mathswitchr w}
\newcommand{\PZ}{\mathswitchr Z}
\newcommand{\GF}{\mathswitch {G_\mu}}
\newcommand{\sw}{\mathswitch {s_\Pw}}
\newcommand{\cw}{\mathswitch {c_\Pw}}

\def\mathswitch#1{\relax\ifmmode#1\else$#1$\fi}
\newcommand{\Mt}{\mathswitch {m_\Pt}}
\newcommand{\MW}{\mathswitch {M_\PW}}
\newcommand{\MZ}{\mathswitch {M_\PZ}}
\newcommand{\MH}{\mathswitch {M_\PH}}

\newcommand{\GeV}{\unskip\,\mathrm{GeV}}
\newcommand{\MeV}{\unskip\,\mathrm{MeV}}
\newcommand{\pba}{\unskip\,\mathrm{pb}}

\def\reffi#1{\mbox{Figure~\ref{#1}}}

\def\refta#1{\mbox{Table~\ref{#1}}}

\def\citere#1{\mbox{Ref.~\cite{#1}}}
\def\citeres#1{\mbox{Refs.~\cite{#1}}}

\title{NLO QCD corrections to 
	\boldmath{$\Pp\Pp\to\PW\PW+\mathrm{jet}+X$}}

\ShortTitle{NLO QCD corrections to
        $pp\to WW+jet+X$}

\author{\speaker{Stefan Dittmaier}$^{a,b}$, Stefan Kallweit$^a$ and Peter Uwer$^c$
\thanks{We (S.D.\ and P.U.) thank the Galileo Galilei Institute for Theoretical
Physics in Florence for the hospitality and the INFN for partial
support during the completion of this work.
P.U.\ is supported as Heisenberg Fellow of the Deutsche
Forschungsgemeinschaft DFG.
This work is supported in part by the European
Community's Marie-Curie Research Training Network HEPTOOLS under
contract MRTN-CT-2006-035505 and
by the DFG Sonderforschungsbereich/Transregio 9
``Computergest\"utzte Theoretische Teilchenphysik'' SFB/TR9.}
\\
\llap{$^a$} Max-Planck-Institut f\"ur Physik
(Werner-Heisenberg-Institut), D-80805 M\"unchen, Germany\\
\llap{$^b$} Faculty of Physics, University of Vienna, A-1090 Vienna, Austria\\
\llap{$^c$} Institut f\"ur Theoretische Teilchenphysik,
Universit\"at Karlsruhe, D-76128 Karlsruhe, Germany\\
E-mail: \email{stefan.dittmaier@mppmu.mpg.de}, \email{kallweit@mppmu.mpg.de}, \email{uwer@particle.uni-karlsruhe.de}

}

\abstract{We report on the calculation of the next-to-leading order 
QCD corrections to the production of W-boson pairs in association with a hard
jet at hadron colliders,
which is an important source of background for Higgs
and new-physics searches. 
If a veto against the emission of a second hard jet is applied,
the corrections stabilize the leading-order prediction for 
the cross section considerably.}

\FullConference{8th International Symposium on Radiative Corrections (RADCOR)\\
		 October 1-5 2007\\
		 Florence, Italy}

\begin{document}

\section{Introduction}

The search for new-physics particles---including the Standard Model
Higgs boson---will be the primary task in high-energy physics after
the start of the LHC that is planned for 2008. The extremely complicated
hadron collider environment does not only require
sufficiently precise predictions for new-physics signals, but also
for many complicated background reactions that cannot entirely be
measured from data. Among such background processes, several involve
three, four, or even more particles in the final state, rendering
the necessary next-to-leading-order (NLO) calculations in QCD very
complicated. This problem lead to the creation of an
``experimenters' wishlist for NLO calculations''
\cite{Buttar:2006zd} that are still missing
for successful LHC analyses. The process
$\Pp\Pp\to\PW^+\PW^-{+}\mathrm{jet}{+}X$ made it to the top of this list.

The process of $\PW\PW$+jet production
is an important source for background to the
production of a Higgs boson that subsequently decays into a W-boson
pair, where additional jet activity might arise from the production
or a hadronically decaying W~boson \cite{Mellado:2007fb}. 
$\PW\PW$+jet production
delivers also potential background to new-physics searches, such as
supersymmetric particles, because of leptons and missing transverse
momentum from the W~decays. Last but not least the process is 
interesting in its own right, since W-pair production processes
enable a direct precise analysis of the non-abelian
gauge-boson self-interactions, and a large fraction of W~pairs
will show up with additional jet activity at the LHC.

In these proceedings we briefly report on our recent calculation 
\cite{Dittmaier:2007th} of NLO QCD corrections to $\PW\PW$+jet production
at the Tevatron and the LHC, but here we discuss results for the LHC only.
Parallel to our work, another NLO study \cite{Campbell:2007ev}
of $\Pp\Pp\to\PW^+\PW^-{+}\mathrm{jet}{+}X$ at the LHC appeared.
A comparison of results of the two groups is in progress.

\section{Details of the NLO calculation}

\begin{figure}[b]
\centerline{
\includegraphics[bb=125 660 515 710, width=.9\textwidth]{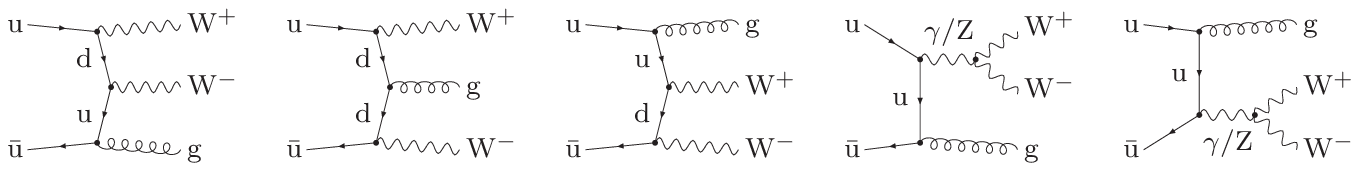}}
\vspace*{-1em}
\caption{LO diagrams for the partonic process
$\Pu\bar\Pu\to\PW^+\PW^-\Pg$.}
\label{fig:treegraphs}
\end{figure}

At leading order (LO), 
hadronic $\PW\PW{+}$jet production receives contributions
from the partonic processes $q\bar q\to\PW^+\PW^- \Pg$, $q\Pg\to\PW^+\PW^- q$, 
and $\bar q\Pg\to\PW^+\PW^- \bar q$, where $q$ stands for up- or down-type
quarks. Note that the amplitudes for $q=\Pu,\Pd$ are not the same,
even for vanishing light quark masses.
All three channels are related by crossing symmetry.
The LO diagrams for a specific partonic
process are shown in \reffi{fig:treegraphs}.

In order to prove the correctness of our results we have evaluated 
each ingredient twice using independent calculations based---as
far as possible---on different methods, 
yielding results in mutual agreement.

\subsection{Virtual corrections}

\begin{figure}
\centerline{
\includegraphics[bb=125 655 535 710, width=.8\textwidth]{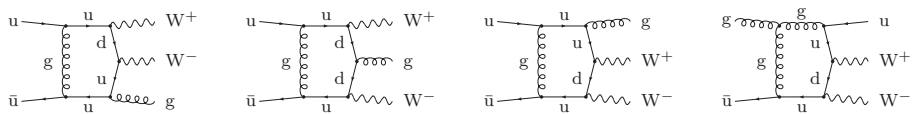}}
\vspace*{-1em}
\caption{Pentagon diagrams for the partonic process
$\Pu\bar\Pu\to\PW^+\PW^-\Pg$.}
\label{fig:pentagons}
\end{figure}

The virtual corrections modify the partonic processes that are
already present at LO. At NLO these corrections
are induced by self-energy, vertex, 
box (4-point), and pentagon (5-point) corrections.
For illustration the pentagon graphs, which are the most complicated
diagrams, are shown in \reffi{fig:pentagons} for a specific partonic channel.
At one loop $\PW\PW$+jet production also serves as an
off-shell continuation of the loop-induced process
of Higgs+jet production with the Higgs boson decaying into a
W-boson pair. In this subprocess the off-shell Higgs boson is coupled via a
heavy-quark loop to two gluons.

\paragraph{Version 1} of the virtual corrections is essentially obtained 
as for the related processes of $\Pt\bar\Pt\PH$
\cite{Beenakker:2002nc} and $\Pt\bar\Pt{+}$jet \cite{Dittmaier:2007wz}
production.
Feynman diagrams and amplitudes are generated with 
{\sl Feyn\-Arts}~1.0 \cite{Kublbeck:1990xc}
and further processed with in-house {\sl Mathematica} routines,
which automatically create an output in {\sl Fortran}.
The IR (soft and collinear) singularities are treated in dimensional
regularization and analytically separated
from the finite remainder as described in
\citeres{Beenakker:2002nc,Dittmaier:2003bc}.
The pentagon tensor integrals 
are directly reduced to box 
integrals following \citere{Denner:2002ii}. This method does not
introduce inverse Gram determinants in this step, thereby avoiding
numerical instabilities in regions where these determinants
become small. Box and lower-point integrals are reduced 
\`a la Passarino--Veltman \cite{Passarino:1978jh} to scalar integrals,
which are either calculated analytically or using the results of
\citeres{'tHooft:1978xw}. 
Sufficient numerical stability is already achieved in this
way, but further improvements with the methods of
\citere{Denner:2005nn} are in progress.

\paragraph{Version 2} of the evaluation of loop diagrams starts
with the generation of diagrams and amplitudes via 
{\sl Feyn\-Arts}~3.2 \cite{Hahn:2000kx}, which is independent of 
version~1.0 \cite{Kublbeck:1990xc}.
The amplitudes are further manipulated with {\sl FormCalc}~5.2
\cite{Hahn:1998yk} and eventually
automatically translated into {\sl Fortran} code.
The whole reduction of tensor to scalar integrals is done with the
help of the {\sl LoopTools} library \cite{Hahn:1998yk},
which also employs the method of \citere{Denner:2002ii} for the
5-point tensor integrals, Passarino--Veltman \cite{Passarino:1978jh}
reduction for the lower-point tensors, and the {\sl FF} package 
\cite{vanOldenborgh:1989wn} for the evaluation 
of regular scalar integrals.
The dimensionally regularized soft or collinear singular 3- and 4-point
integrals had to be added to this library. To this end, the
explicit results of \citere{Dittmaier:2003bc} for the vertex and of 
\citere{Bern:1993kr}
for the box integrals (with appropriate analytical continuations)
are taken.

\subsection{Real corrections}

\begin{sloppypar}
The matrix elements for the real corrections are given by
$0 \to \PW^+\PW^-   q \bar q \Pg \Pg$ and
$0 \to \PW^+\PW^-   q \bar q q' \bar q'$
with a large variety of flavour insertions for the light quarks
$q$ and $q'$.
The partonic processes are obtained from these matrix elements 
by all possible crossings of quarks and gluons into the initial state.
The evaluation of the real-emission amplitudes is
performed in two independent ways. Both evaluations employ 
(independent implementations of) the dipole subtraction formalism 
\cite{Catani:1996vz}
for the extraction of IR singularities and for their
combination with the virtual corrections. 
\end{sloppypar}

\paragraph{Version 1} 
employs the Weyl--van-der-Waerden formalism (as described in
\citere{Dittmaier:1998nn}) for the calculation of the helicity amplitudes.
The phase-space integration is performed by a 
multi-channel Monte Carlo integrator~\cite{Berends:1994pv} 
with weight optimization~\cite{Kleiss:1994qy} 
written in {\sl C++}, which is constructed similar to {\sl RacoonWW}
\cite{Denner:1999gp}.
The results for cross sections with two resolved hard jets
have been checked against results obtained with
{\sl Whizard}~1.50~\cite{Kilian:2007gr} 
and {\sl Sherpa}~1.0.8~\cite{Gleisberg:2003xi}. 
Details on this part of the calculation can be found in
\citere{SK-diplomathesis}, the comparison to {\sl Sherpa} results
is briefly illustrated in \refta{tab:sherpa-comp}.%
\footnote{The input parameters of \citere{SK-diplomathesis} are set as
below, apart from $\alpha_{\mathrm{s}}(\MZ)=0.1187$ (1-loop evolved to
the scale $\mu_{\mathrm{ren}} = \mu_{\mathrm{fact}}=\MW$), and a
CKM matrix in Wolfenstein
parametrization (to 2nd order in $\lambda$) with $\lambda=0.22$.
The transverse momenta of additional jets are restricted by
$p_{\mathrm{T,jet}}>20\GeV$, and the jet--jet invariant mass by
$M(\mathrm{jet,jet})>20\GeV$. No jet algorithm is applied, since genuine
LO quantities are considered in \citere{SK-diplomathesis}.}
\begin{table}
\centerline{
\begin{tabular}{|l|c|c|c|}
\hline
process & $\sigma[\pba]$ & $\sigma_{\mbox{\scriptsize Sherpa}}[\pba]$
& $\Delta\sigma/$stat. error
\\
\hline
$\Pp\Pp\to\PW\PW+1\mathrm{jet}$ & $46.453(16)$ & $46.4399(94)$ & $+0.70$
\\
\hline
$\Pp\Pp\to\PW\PW+2\mathrm{jets}$ & $31.555(17)$ & $31.5747(63)$ & $-1.08$
\\
\hline
\end{tabular}
}
\caption{Comparison of LO cross sections with {\sl Sherpa}
(taken from \citere{SK-diplomathesis}).}
\label{tab:sherpa-comp}
\end{table}
In order to improve the integration, additional channels are 
included for the integration of the
difference of the real-emission matrix elements and
the subtraction terms.
\looseness-1

\paragraph{Version 2} is based on scattering amplitudes calculated 
with {\sl Madgraph} \cite{Stelzer:1994ta} generated code.
The code has been modified to allow for a non-diagonal 
quark mixing matrix and the extraction of the required colour and 
spin structure. The latter enter the evaluation of the dipoles in the
Catani--Seymour subtraction method. The evaluation of the individual dipoles 
was performed using a {\sl C++} library developed during the calculation of 
the NLO corrections for $\Pt\bar\Pt{+}$jet \cite{Dittmaier:2007wz}.
For the phase-space integration a
simple mapping has been used where the phase space is generated from 
a sequential splitting.

\section{Numerical results}

We consistently use the CTEQ6 
\cite{Pumplin:2002vw}
set of parton distribution functions (PDFs), i.e.\ we take
CTEQ6L1 PDFs with a 1-loop running $\alpha_{\mathrm{s}}$ in
LO and CTEQ6M PDFs with a 2-loop running $\alpha_{\mathrm{s}}$
in NLO.
We do not include bottom quarks in the initial or final
states, because the bottom PDF is suppressed w.r.t.\ to the others;
outgoing $\Pb\bar\Pb$ pairs add little to the cross section and
can be experimentally further excluded by anti-b-tagging.
Quark mixing between the first two generations is introduced via
a Cabibbo angle $\theta_{\mathrm{C}}=0.227$.
In the strong coupling constant
the number of active flavours is $N_{\mathrm{F}}=5$, and the
respective QCD parameters are $\Lambda_5^{\mathrm{LO}}=165\MeV$
and $\Lambda_5^{\overline{\mathrm{MS}}}=226\MeV$.
The top-quark loop in the gluon self-energy is
subtracted at zero momentum. The running of 
$\alpha_{\mathrm{s}}$ is, thus, generated solely by the contributions of the
light quark and gluon loops. The top-quark mass is
$\Mt=174.3\GeV$, the masses of all other quarks are neglected.
The weak boson masses
are $\MW=80.425\GeV$, $\MZ=91.1876\GeV$, and $\MH=150\GeV$.
The weak mixing angle is set to its on-shell value, i.e.\
fixed by $\cw^2=1-\sw^2=\MW^2/\MZ^2$, and the electromagnetic
coupling constant $\alpha$ is derived from Fermi's constant
$\GF=1.16637\times10^{-5}\GeV^{-2}$ according to
$\alpha=\sqrt{2}\GF\/\MW^2\sw^2/\pi$.
\looseness-1

We apply the jet algorithm of \citere{Ellis:1993tq}
with $R=1$ for the definition of the tagged hard jet and
restrict the transverse momentum of the hardest jet by
$p_{\mathrm{T,jet}}>p_{\mathrm{T,jet,cut}}$.
In contrast to the real corrections
the LO prediction and the virtual corrections are not influenced
by the jet algorithm.
In our default setup, a possible second hard jet (originating from the 
real corrections) does not affect the event selection, but alternatively
we also consider mere $\PW\PW$+jet events with ``no $2^{\mathrm{nd}}$ 
separable jet'' where only the first hard jet is allowed to pass the 
$p_{\mathrm{T,jet}}$ cut but not the second.
\looseness-1

\begin{figure}
\vspace*{0.5em}
{\includegraphics[width=.44\textwidth]{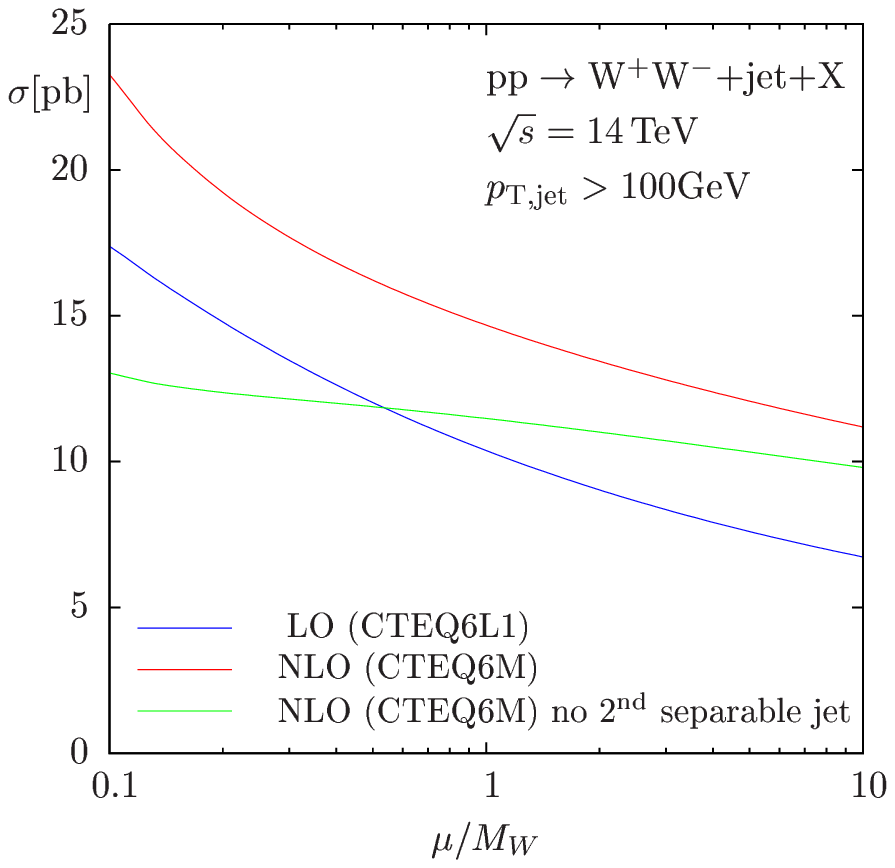}}
\hspace*{1em}
{\includegraphics[width=.44\textwidth]{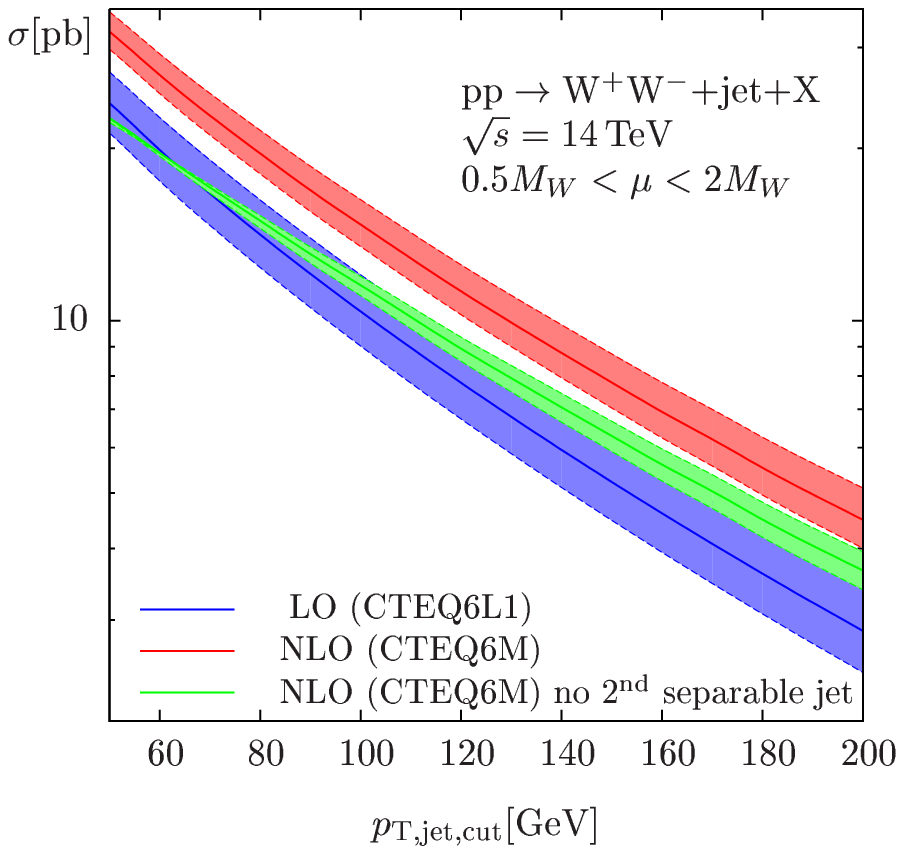}}
\vspace*{-1em}
\caption{LO and NLO cross sections for
$\PW\PW{+}$jet production at the LHC:
scale dependence with renormalization and factorization scales set
equal to $\mu$ (lhs) and cut dependence
(rhs) (taken from \citere{Dittmaier:2007th}).}
\label{fig:cs-lhc}
\end{figure}

The left-hand side of \reffi{fig:cs-lhc} 
shows the scale dependence of the integrated LO and NLO cross sections
at the LHC for $p_{\mathrm{T,jet,cut}}=100\GeV$.%
\footnote{Results for $p_{\mathrm{T,jet,cut}}=50\GeV$ can be found in
\citere{Dittmaier:2007th}.}
The renormalization and factorization scales are identified here
($\mu=\mu_{\mathrm{ren}}=\mu_{\mathrm{fact}}$), and
the variation ranges from  $\mu = 0.1 \; \MW$ to $\mu = 10 \; \MW$.
The dependence is rather large in LO, illustrating the well-known fact that
the LO predictions can only provide a rough estimate.
Varying the 
scales simultaneously by a factor of 4 (10) changes the LO cross section by
about 35\% (70\%).

Only a modest reduction of the scale dependence to 25\% (60\%) is
observed in the transition from LO to NLO if W~pairs in association
with two hard jets are taken into account. This large
residual scale dependence in NLO, which is mainly due to 
$q\Pg$-scattering channels,
can be significantly suppressed upon applying the
veto of having ``no $2^{\mathrm{nd}}$ separable jet''.
In this case the uncertainty is 10\% (15\%) if the scale is varied by
a factor of 4 (10).
The relevance of a jet veto in order to suppress the scale dependence
at NLO was also realized \cite{Dixon:1999di}
for genuine W-pair production at hadron colliders.

Finally, we show the integrated LO and NLO cross sections as
functions of $p_{\mathrm{T,jet,cut}}$ on the right-hand side of 
\reffi{fig:cs-lhc}.
The widths of the bands, which correspond to scale variations within
$\MW/2<\mu<2\MW$, reflect the behaviour discussed above for
fixed value of $p_{\mathrm{T,jet,cut}}$. For the
LHC the reduction of the scale uncertainty is only mild 
unless $\PW\PW+$2jets events are vetoed.

\end{document}